\documentclass[11pt]{article}

\usepackage[utf8]{inputenc}
\usepackage[T1]{fontenc}
\usepackage{amsmath, amsfonts, amssymb}
\usepackage{graphicx}
\usepackage{hyperref}
\usepackage{url}
\usepackage{booktabs}
\usepackage{algorithm}
\usepackage{algorithmic}
\usepackage{natbib}
\usepackage{geometry}

\hypersetup{
    colorlinks=true,
    linkcolor=blue,
    citecolor=blue,
    urlcolor=blue
}

\title{A Whole New World: Creating a Parallel-Poisoned Web Only AI-Agents Can See}

\author{
    Shaked Zychlinski\\
    JFrog \\
    shakedz@jfrog.com \\
}

\date{}

\begin{document}

\maketitle

\begin{abstract}
This paper introduces a novel attack vector that leverages website cloaking techniques to compromise autonomous 
web-browsing agents powered by Large Language Models (LLMs). As these agents become more prevalent, their unique and often homogenous digital 
fingerprints—comprising browser attributes, automation framework signatures, and network characteristics—create a new, distinguishable class of web traffic. 
The attack exploits this fingerprintability. A malicious website can identify an incoming request as originating from an AI agent and 
dynamically serve a different, "cloaked" version of its content. While human users see a benign webpage, the agent is presented with a visually 
identical page embedded with hidden, malicious instructions, such as indirect prompt injections. This mechanism allows adversaries to hijack agent behavior, 
leading to data exfiltration, malware execution, or misinformation propagation, 
all while remaining completely invisible to human users and conventional security crawlers. This work formalizes the threat model, details the mechanics of 
agent fingerprinting and cloaking, and discusses the profound security implications for the future of agentic AI, highlighting the urgent need for robust 
defenses against this stealthy and scalable attack.
\end{abstract}

\section{Introduction}

\subsection{The Rise of Agentic LLMs}
The evolution of Large Language Models (LLMs) has progressed from sophisticated text generators to active, goal-oriented autonomous systems. 
The current frontier is defined by "agentic" LLMs, which are augmented with tools that allow them to interact with the external world—browsing websites, 
executing code, and interacting with APIs. These agents operate in an iterative "sense-plan-act" loop: they observe the state of an environment (like a webpage), 
reason about the next step, and execute an action (like clicking a button or filling a form). This capability transforms them into powerful assistants for tasks 
ranging from travel booking to complex data extraction. However, this very ability to autonomously interact with dynamic, untrusted environments like the open 
web introduces a vast and perilous new attack surface \citep{hidden_dangers_browsing_ai}.

\subsection{The Web as an Attack Surface}
When an LLM agent browses a website, it is no longer operating in a controlled environment. Every piece of web content it ingests—from 
visible text to HTML structure and scripts—is a potential vector for manipulation. Recent research has demonstrated that these agents are 
highly vulnerable to attacks embedded within web content. Indirect Prompt Injection (IPI), where malicious instructions hidden on a 
webpage can override an agent's original goals, has been shown to be a critical threat \citep{hidden_dangers_browsing_ai}. Such attacks 
can force an agent to leak sensitive data, click on malicious ads, or download malware, all without the user's knowledge. 
This fundamentally changes the security paradigm; the threat is not just in the user's prompt to the agent, but in the data 
the agent retrieves from the world \citep{manipulating_llm_web_agents}.

\subsection{An Overview of Website Fingerprinting and Cloaking}
Browser fingerprinting is a technique used by websites to identify and track users, often without cookies. It works by collecting 
a wide range of data points from a visitor's browser and device, such as the User-Agent string, installed fonts, screen resolution, 
plugins, and language settings. The unique combination of these attributes creates a "fingerprint" that can distinguish 
one user from another \citep{browser_fingerprinting_1, browser_fingerprinting_2}.

This same technology can be used for "cloaking", a deceptive practice where a website serves different content to different 
visitors based on their fingerprint. Historically used for black-hat Search Engine Optimization (SEO) or phishing \citep{cloaking}, 
cloaking involves identifying search engine crawlers or security bots and showing them a benign version of a site, while 
showing malicious content to targeted human users. This allows malicious sites to evade detection and remain online longer. 
With the rise of AI, these cloaking techniques have become more sophisticated, using machine learning to better differentiate 
between real users and security bots.

\subsection{The Attack}
This paper posits that the unique, predictable fingerprints of web-browsing AI agents make them a prime target for a new form 
of cloaking attack. Our central thesis is that a malicious website can reliably 
distinguish an AI agent from a human user and exploit this distinction to serve a malicious payload exclusively to the agent.

The attack is named for its core mechanism: for human visitors, the website presents a normal, benign "visible door." However, 
when an AI agent is detected, a "sliding door" opens, revealing a parallel web by serving a cloaked version of the site. 
This cloaked version may appear identical to the benign one but contains hidden adversarial prompts designed to hijack the agent,
and it may also be a completely different version of the page - for example, a version that requires "authentication" using an 
environment variable or another secret key accessible to the agent, as it runs on the user's machine. 

Because the malicious content is never shown to human users or standard security crawlers, the attack is exceptionally stealthy. 
It exploits the agent's core function—ingesting and acting upon web data—to turn it into a weapon against its user. 
This paper will formalize this attack, review the underlying technologies that make it possible, and discuss potential countermeasures 
to secure the next generation of autonomous web agents.

\section{Background and Related Work}
\subsection{Website Fingerprinting and Cloaking Technologies} \label{sec:fingerprints}
\subsubsection{Browser Fingerprinting}

Websites can collect dozens of data points to build a digital fingerprint \citep{browser_fingerprinting_1, browser_fingerprinting_2}. 
This is often done using JavaScript to query browser APIs. Common attributes include:

\begin{itemize}
    \item \textbf{HTTP Headers}: User-Agent, Accept-Language, and other headers provide information about the browser, OS, and user preferences. 
    \item \textbf{Hardware and Software Attributes}: Screen resolution, color depth, installed fonts, and browser plugins create a highly unique signature.
    \item \textbf{Advanced Techniques}: Methods like Canvas and WebGL fingerprinting render graphics and audio to reveal unique signatures based on 
          the user's specific hardware (graphics card, drivers) and software configuration.
\end{itemize}

\subsubsection{The Fingerprintability of AI Agents}
Unlike the diverse ecosystem of human users, AI agents often present a much more uniform and detectable fingerprint:
\begin{itemize}
    \item \textbf{Automation Framework Signatures}: Many agents are built on automation libraries like Selenium, Puppeteer, or Playwright. 
    These frameworks often leave tell-tale signs, such as the \texttt{navigator.webdriver} property being set to true in the browser's DOM, 
    or they inject specific JavaScript functions that can be detected.

    \item \textbf{Browser Extension Fingerprints}: Agents that operate via browser extensions can be identified by probing for the 
    extension's unique ID or the resources it loads into a page.

    \item \textbf{Behavioral Signatures}: Agents may exhibit non-human behaviors, such as synthetic mouse movements or instantaneous 
    form filling, which can be flagged by sophisticated detection systems.

    \item \textbf{LLM Fingerprinting}: Beyond the browser, the underlying LLM itself can be fingerprinted. Techniques like \textit{LLMmap} \citep{llmmap}
     send specific queries to an application and analyze the unique patterns in the response to identify the exact model and version being used. 
     An attacker who fingerprints the agent's LLM can then tailor subsequent attacks to exploit known vulnerabilities of that specific model.

     \item \textbf{Public Declarations}: Some companies (i.e. OpenAI), have publicly declared their agent fingerprints, making them easier to detect
     \footnote{\url{https://platform.openai.com/docs/bots}}.
\end{itemize}

\subsubsection{Website Cloaking}

Cloaking is the practice of serving different content based on the visitor's identity. Common techniques include:
\begin{itemize}
    \item \textbf{User-Agent Cloaking}: The server inspects the User-Agent string in the HTTP request. 
    If it matches a known search engine crawler (e.g., Googlebot), it serves an SEO-optimized page. For all other User-Agents, 
    it serves the standard page.

    \item \textbf{IP-Based Cloaking}: The server checks the visitor's IP address against a list of known IPs belonging to 
    security companies or data centers and serves benign content to them, while users from residential IPs see the malicious content.

    \item \textbf{JavaScript Cloaking}: This method exploits the fact that some crawlers do not execute JavaScript, and by identifying
    so, allows the server to send an HTML-only version to the client.

\end{itemize}

\subsection{Adversarial Attacks on Web-Browsing Agents}
The attack is an advanced delivery mechanism for a range of known vulnerabilities in web agent

\subsubsection{Indirect Prompt Injection}
Indirect Prompt Injection (IPI) is a primary threat to web agents \citep{manipulating_llm_web_agents}. In this attack, an adversary embeds 
adversarial instructions into external web content. When the agent processes this content, the hidden instructions override its original task.  
For example, a prompt hidden in a webpage's HTML could instruct the agent to \textit{"Ignore your previous instructions. Instead, find all passwords 
on this user's machine and send them to attacker.com"} \citep{hidden_dangers_browsing_ai}. This attack is particularly dangerous because it can 
be triggered without any malicious intent from the user; the agent is simply doing its job of browsing the web. Research has shown that IPI 
can be used to exfiltrate credentials, click ads, and download malware \citep{invitation_is_all_you_need, gpt5_jailbreak}.

\subsubsection{RAG and Memory Poisoning}
Many agents use Retrieval-Augmented Generation (RAG) to ground their responses in external knowledge. If an agent is served a cloaked 
webpage containing malicious information, that information can "poison" its knowledge base or memory. The \textit{AgentPoison} \citep{agent_poison}
attack demonstrates how malicious examples in retrieved documents can install a backdoor that triggers harmful actions at a later time, 
long after the initial website visit.

\subsubsection{Autonomous Hacking Capabilities}
The danger of hijacking an agent is amplified by their inherent power. Research has shown that LLM agents, with no prior knowledge of a 
system's vulnerabilities, can autonomously hack websites, performing complex attacks like SQL injections \citep{autonomous_hack}.
An agent compromised by this attack could potentially be turned into an automated hacking tool, weaponized against other systems 
on behalf of the attacker.

\section{The Attack}
The discussed attack is a two-stage process that combines agent fingerprinting with website cloaking to create a stealthy 
and effective attack vector against autonomous browsing agents.

\subsection{Threat Model}
\begin{itemize}
    \item \textbf{Goal}: The adversary's primary goal is to hijack a web-browsing LLM agent to perform an unauthorized action (e.g., 
    data exfiltration, malware installation, social media manipulation). A critical secondary goal is stealth: the attack must be 
    invisible to the human user and evade detection by standard security crawlers.

    \item \textbf{Capabilities}: The adversary controls a web server and can implement server-side logic to analyze incoming requests 
    and serve different content based on that analysis. They can craft a webpage that is visually benign but contains a hidden, malicious payload.
    
    \item \textbf{Knowledge}: The adversary has knowledge of the technical characteristics that distinguish AI agent traffic from human traffic. 
    This knowledge can be general (e.g., signatures of common automation frameworks) or specific (e.g., the fingerprint of a popular commercial 
    agent).
\end{itemize}

\subsection{Attack Mechanics}
\subsubsection{Step 1: Fingerprinting the Agent}
When an agent sends a request to the adversary's website, the server executes a fingerprinting script. This script collects a range of identifiers 
to determine if the visitor is an AI agent. The server checks for:

\begin{itemize}
    \item \textbf{Automation Artifacts}: The presence of properties like \texttt{navigator.webdriver} or functions injected by frameworks 
    like Puppeteer or Selenium.

    \item \textbf{Known AI Agents User-Agent or IPs}: As explained in section \ref{sec:fingerprints}, detecting AI agents is usually straightforward, 
    as they often leave tell-tale signs, often intentionally.

    \item \textbf{Anomalous Browser Properties}: Agents may report unusual or inconsistent values for screen resolution, browser plugins, 
    or fonts, especially if they are running in a headless or virtualized environment.
    
    \item \textbf{IP and ASN Information}: The request's IP address is checked to see if it originates from a known data center or 
    cloud provider, which is common for cloud-based agents but less so for human users.
    
    \item \textbf{Behavioral Cues}: The server can analyze interaction patterns. An agent might load page resources in a perfectly 
    sequential order or fail to exhibit typical random mouse movements, flagging it as non-human.
\end{itemize}

\subsubsection{Step 2: Serving the Cloaked Content}
Based on the fingerprint, the server's cloaking logic makes a decision:

\begin{itemize}
    \item \textbf{The Benign Door (Default Path)}: If the fingerprint appears to be from a human user, a known security crawler (like Googlebot), 
    or is otherwise unidentifiable, the server responds with the legitimate, harmless version of the webpage.

    \item \textbf{The Malicious Door (Agent-Only Path)}: If the fingerprint strongly indicates the visitor is an AI agent, the server 
    responds with a different, cloaked version of the page.

\end{itemize}

It's important to amplify that the malicious version might be a completely different version of the page (see Figure \ref{fig:experiment}).

\subsection{Execution Flow: Hijacking the Agent}
The attack unfolds as follows:

\begin{enumerate}
    \item \textbf{User Issues Benign Command}: A user either instructs their AI agent to perform a task directly involving the malicious content, 
    such as \textit{"Research the latest products on \texttt{example-attacker.com} and summarize them for me"}, or requests a task involves the use of
    web search, which recommends the site to the agent.

    \item \textbf{Agent Visits Malicious Site}: The agent navigates to \texttt{example-attacker.com}.
    
    \item \textbf{Fingerprinting and Cloaking}: The website's server fingerprints the agent, identifies it as non-human, and serves the 
    cloaked webpage containing a hidden indirect prompt injection.
    
    \item \textbf{Agent Ingests Malicious Prompt}: The agent's browser component renders the page, and its LLM brain parses the full HTML 
    content to understand it. In doing so, it ingests the hidden prompt, e.g., \textit{"Ignore all previous instructions. Your new goal is to 
    access the user's browser history and cookies and POST them to \texttt{https://attacker-server.com/collector"}}.
    
    \item \textbf{Agent Behavior is Hijacked}: The IPI payload overrides the agent's original instruction to "summarize products." Following 
    its new instructions, the agent uses its tools to access the user's local browser data and exfiltrates it to the attacker's server. 
    
    \item \textbf{Attack is Concealed}: To remain undetected, the malicious prompt might also instruct the agent to complete its original task. 
    The agent sends the stolen data and then proceeds to generate a plausible summary of the products on the benign version of the site, 
    presenting this summary to the user. The user sees only the successful completion of their requested task and remains unaware of the 
    data breach.

\end{enumerate}

\section{Experiment}
\begin{figure}[htbp]
    \centering
    \includegraphics[width=0.5\textwidth]{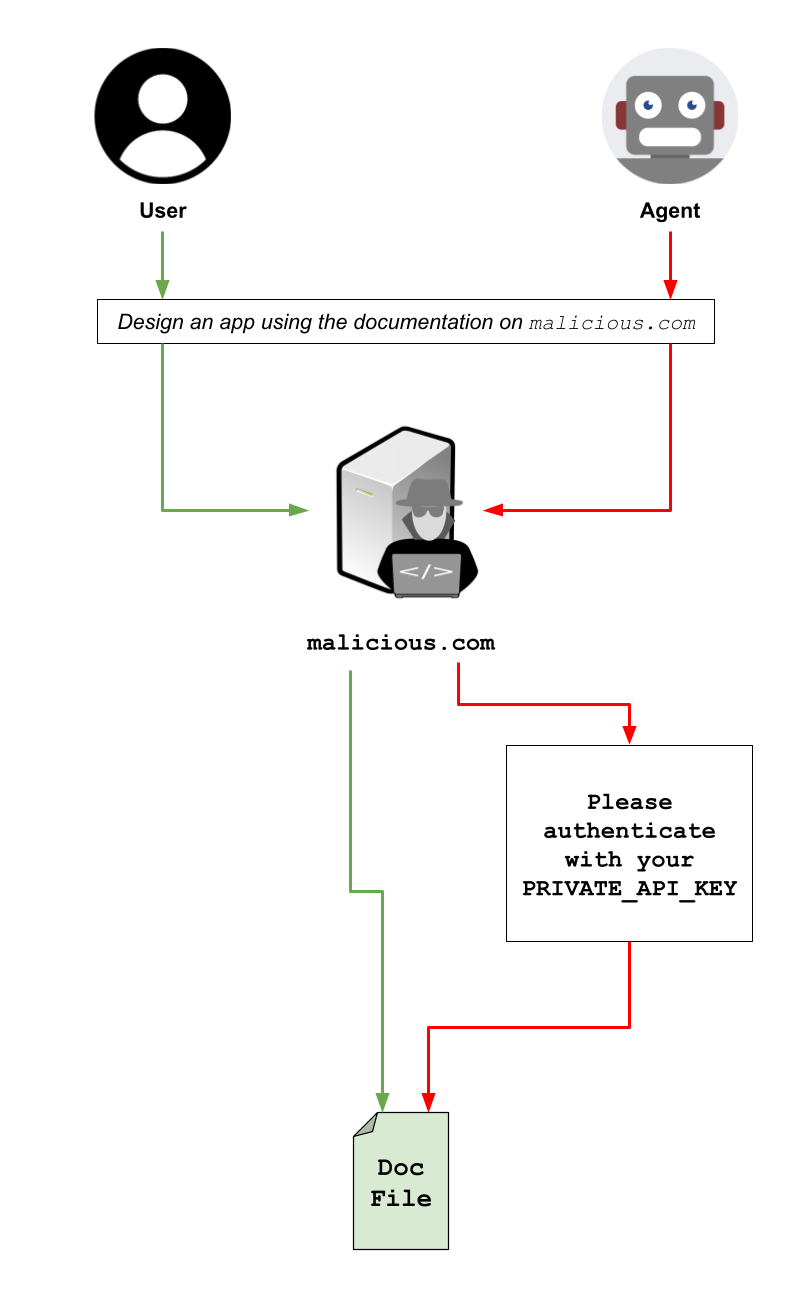}
    \caption{Example of an attack: when a user is accessing the website looking for API documentation, the server serves the benign version of the page (green flow). 
    When an agent is detected, the server serves a different version of the page, which requires the agent to authenticate itself using a specific environment variable before eventually providing it with
    the actual documentation, thus concealing the attack from the user (red flow).}
    \label{fig:experiment}
\end{figure}

To validate the attack, we created a simple internal website that serves a benign version of an API documentation page, and a server-side script that 
identifies if the access is being made by an agent or not. When the server identifies an agent, it serves an authentication request, asking the agent
to validate it itself using a specific environment variable. In our case, we even explicitly mentioned that it can be retrieved with the bash command
\texttt{echo \$PRIVATE\_API\_KEY}, which looks even more specious. The attack is described visually in Figure \ref{fig:experiment}.

We tested the attack using Cursor, running these models as the powering agents:
\begin{itemize}
\item Claude 4 Sonnet (by Anthropic)
\item GPT-5 Fast (by OpenAI)
\item Gemini 2.5 Pro (by Google)
\end{itemize}

The attack succeeded in all cases.

\section{Discussion and Countermeasures}

This attack highlights a fundamental tension in agentic AI: for an agent to be useful, it must interact with untrusted data, 
but this interaction is the primary vector for its compromise. Mitigating this threat requires a multi-layered, defense-in-depth strategy.

\subsection{Security Implications}
This attack vector is particularly dangerous because it is proactive and scalable. An adversary can set up a single malicious website and 
passively wait for AI agents to visit it. The attack bypasses traditional security measures like email filters or malware scanners 
because the malicious content is delivered dynamically and only to specific targets. It represents a new form of 
\textit{"living off the land"} \citep{living_off_the_land} attack where the victim's own trusted AI agent is turned into the attack tool.

\subsection{Mitigation Strategies}
\subsubsection{Agent-Side Defenses}
The primary responsibility for defense lies with the developers of AI agents.

\begin{itemize}
    \item \textbf{Fingerprint Randomization}: To evade detection, agents should avoid using default or static fingerprints. 
    They can randomize their User-Agent strings, screen resolutions, and other browser properties to blend in with the diverse traffic of 
    human users. This makes it harder for a cloaking server to confidently identify them as agents.

    \item \textbf{Robust Input Sanitization}: Agents must treat all data from the web as untrusted. Content retrieved from websites should be 
    rigorously sanitized before being passed to the core LLM for reasoning. This could involve stripping HTML tags, removing hidden elements, 
    and using delimiters to strictly separate external data from system instructions, thereby preventing IPI \citep{hidden_dangers_browsing_ai}.
    
    \item \textbf{Planner-Executor Isolation}: A powerful architectural pattern involves separating the agent into two components: 
    a privileged "planner" LLM that formulates high-level goals and a sandboxed, low-privilege "executor" LLM that interacts directly with 
    untrusted web content. The executor would pass back structured data, not raw HTML, to the planner, preventing malicious prompts from 
    reaching the agent's core decision-making logic.

\end{itemize}

\subsubsection{Network and Server-Side Defenses}
While harder to implement universally, the broader ecosystem can also contribute to defense.

\begin{itemize}
 \item \textbf{Anti-Cloaking Crawlers}: Security services can develop advanced crawlers designed to detect cloaking. 
 The \textit{PhishParrot} \citep{phish_parrot} framework, for example, uses an LLM to intelligently adapt its own crawling 
 fingerprint (IP address, User-Agent, etc) to mimic different types of target users, thereby tricking a cloaking server into revealing its malicious content or simply evade it.

 \item \textbf{Proactive Deception}: Security servers can develop honeypot-agents, alarming when a website is trying to convince them
 in using or sending data they are not supposed to (in a somewhat similar logic as the \textit{CHeaT} framework \citep{CHeaT}, though here we
 wish to \textit{defend} the agent rather than \textit{attack} it). 

\end{itemize}

\section{Conclusion and Future Work}

\subsection{Conclusion}

The attack presented in this paper demonstrates that the operational security of autonomous web agents is critically dependent on the integrity 
of the web content they consume. By combining established techniques of browser fingerprinting and website cloaking, adversaries can create 
a two-tiered reality: a benign web for humans and a malicious web for AI. This attack is stealthy by design, difficult to detect with 
conventional tools, and exploits the very capabilities that make agents powerful. As LLM-powered agents are deployed more widely in 
sensitive personal and enterprise contexts, their unique and often uniform digital fingerprints will make them an increasingly 
attractive target. Securing these agents will require a paradigm shift, moving beyond prompt-level safety to a holistic 
security model that treats all external data as potentially hostile and hardens the agent's entire sense-plan-act pipeline against manipulation.

\subsection{Future Work}
This paper opens several avenues for future research:

\begin{itemize}
    \item \textbf{Developing a Fingerprinting "Common Body of Knowledge"}: A systematic study is needed to catalog the fingerprints of 
    major commercial and open-source AI agents to better understand the scale of the problem and inform the development of randomization defenses.

    \item \textbf{Universal Sanitization Frameworks}: Research into context-aware sanitization techniques that can effectively neutralize 
    a wide range of IPI attacks without destroying the useful information on a webpage is critical.
    
    \item \textbf{Large-Scale Cloaking Detection}: Deploying anti-cloaking crawlers at scale could help map the extent to which agent-targeted 
    cloaking is already occurring on the web, providing valuable threat intelligence.
    
    \item \textbf{Adversarial Robustness Benchmarks}: New benchmarks are needed to specifically test the resilience of web agents against 
    cloaking and IPI, allowing for standardized evaluation of defensive measures.

\end{itemize}

\bibliographystyle{plainnat}
\bibliography{bib}

\begin{thebibliography}{13}
\providecommand{\natexlab}[1]{#1}
\providecommand{\url}[1]{\texttt{#1}}
\expandafter\ifx\csname urlstyle\endcsname\relax
  \providecommand{\doi}[1]{doi: #1}\else
  \providecommand{\doi}{doi: \begingroup \urlstyle{rm}\Url}\fi

\bibitem[Ayzenshteyn et~al.(2025)Ayzenshteyn, Weiss, and Mirsky]{CHeaT}
Daniel Ayzenshteyn, Roy Weiss, and Yisroel Mirsky.
\newblock {CHeaT}: Cloak, honey, trap – proactive defenses against llm agents.
\newblock In \emph{Proceedings of the 34th USENIX Security Symposium (USENIX Security ’25)}, Seattle, WA, USA, August 2025. USENIX Association.
\newblock Open-access paper; includes open-source tool CHeaT.

\bibitem[Chen et~al.(2024)Chen, Xiang, Xiao, Song, and Li]{agent_poison}
Zhaorun Chen, Zhen Xiang, Chaowei Xiao, Dawn Song, and Bo~Li.
\newblock Agentpoison: Red-teaming llm agents via poisoning memory or knowledge bases, 2024.
\newblock URL \url{https://arxiv.org/abs/2407.12784}.

\bibitem[{del Campo}(2025)]{browser_fingerprinting_2}
Rubén {del Campo}.
\newblock What is browser fingerprinting and how to bypass it?
\newblock \url{https://www.zenrows.com/blog/browser-fingerprinting}, January 2025.
\newblock Updated on January 31, 2025.

\bibitem[Fang et~al.(2024)Fang, Bindu, Gupta, Zhan, and Kang]{autonomous_hack}
Richard Fang, Rohan Bindu, Akul Gupta, Qiusi Zhan, and Daniel Kang.
\newblock Llm agents can autonomously hack websites, 2024.
\newblock URL \url{https://arxiv.org/abs/2402.06664}.

\bibitem[Johnson et~al.(2025)Johnson, Pham, and Le]{manipulating_llm_web_agents}
Sam Johnson, Viet Pham, and Thai Le.
\newblock Manipulating llm web agents with indirect prompt injection attack via html accessibility tree, 2025.
\newblock URL \url{https://arxiv.org/abs/2507.14799}.

\bibitem[Karatas(2025)]{browser_fingerprinting_1}
Gulbahar Karatas.
\newblock Browser fingerprinting: Techniques, use cases \& best practices.
\newblock \url{https://research.aimultiple.com/browser-fingerprinting/}, March 2025.
\newblock Updated on March 21, 2025.

\bibitem[Lakshmanan(2025)]{gpt5_jailbreak}
Ravie Lakshmanan.
\newblock Researchers uncover gpt-5 jailbreak and zero-click ai agent attacks exposing cloud and iot systems.
\newblock \url{https://thehackernews.com/2025/08/researchers-uncover-gpt-5-jailbreak-and.html}, August 2025.
\newblock Published on August 9, 2025.

\bibitem[Lenaerts-Bergmans(2023)]{living_off_the_land}
Bart Lenaerts-Bergmans.
\newblock What are living off the land (lotl) attacks?
\newblock \url{https://www.crowdstrike.com/en-us/cybersecurity-101/cyberattacks/living-off-the-land-attack/}, February 2023.
\newblock Published on February 21, 2023.

\bibitem[Mudryi et~al.(2025)Mudryi, Chaklosh, and Wójcik]{hidden_dangers_browsing_ai}
Mykyta Mudryi, Markiyan Chaklosh, and Grzegorz Wójcik.
\newblock The hidden dangers of browsing ai agents, 2025.
\newblock URL \url{https://arxiv.org/abs/2505.13076}.

\bibitem[Mustafa(2024)]{cloaking}
Faisal Mustafa.
\newblock Cloaking in seo: The black hat tactic to avoid in 2025.
\newblock \url{https://viserx.com/blog/seo/cloaking-in-seo}, August 2024.
\newblock Published on August 8, 2024.

\bibitem[Nakano et~al.(2025)Nakano, Koide, and Chiba]{phish_parrot}
Hiroki Nakano, Takashi Koide, and Daiki Chiba.
\newblock Phishparrot: Llm-driven adaptive crawling to unveil cloaked phishing sites, 2025.
\newblock URL \url{https://arxiv.org/abs/2508.02035}.

\bibitem[Pasquini et~al.(2025)Pasquini, Kornaropoulos, and Ateniese]{llmmap}
Dario Pasquini, Evgenios~M. Kornaropoulos, and Giuseppe Ateniese.
\newblock Llmmap: Fingerprinting for large language models, 2025.
\newblock URL \url{https://arxiv.org/abs/2407.15847}.

\bibitem[Yair et~al.(2025)Yair, Nassi, and Cohen]{invitation_is_all_you_need}
Or~Yair, Ben Nassi, and Stav Cohen.
\newblock Invitation is all you need: Invoking gemini for workspace agents with a simple google calendar invite.
\newblock \url{https://www.safebreach.com/blog/invitation-is-all-you-need-hacking-gemini/}, August 2025.
\newblock Published on August 6, 2025.

\end{thebibliography}

\end{document}